\documentclass[a4paper]{jpconf}
\usepackage{graphicx}
\begin{document}
\title{Status of the X17 search in Montreal}

%\author{Jacky Mucklow}
\author{G. Azuelos$^{1,2}$, B. Broerman$^3$, D. Bryman$^{2,4}$, W.C. Chen$^1$, H.N. da Luz$^5$, L. Doria$^6$, A. Gupta$^1$, L-A. Hamel$^1$, M. Laurin$^1$, K. Leach$^7$, G. Lefebvre$^1$, J-P. Martin$^1$, A. Robinson$^1$, N. Starinski$^1$, R. Sykora$^5$, D. Tiwari$^8$, U. Wichoski$^{9,10,11}$, V. Zacek$^{1}$}
\address{$^1$ Universit\'e de Montr\'eal, Canada}
\address{$^2$ TRIUMF, Canada}
\address{$^3$ Queen's University, Canada}
\address{$^4$ University of British Columbia, Canada}
\address{$^5$ Czech Technical University, Czech Republic}
\address{$^6$ Johannes Gutenberg University of Mainz, Germany}
\address{$^7$ Colorado School of Mines, USA}
\address{$^8$ University of Regina, Canada}
\address{$^9$ Laurentian University, Canada}
\address{$^{10}$ SNOLAB, Canada}
\address{$^{11}$ Carleton University, Canada}

\ead{georges.azuelos@umontreal.ca}

\begin{abstract}
At the Montreal Tandem accelerator, an experiment is being set up to measure internal pair creation from the decay of nuclear excited states using a multiwire proportional chamber and scintillator bars surrounding it from the DAPHNE experiment. The acceptance covers a solid angle of nearly 4$\pi$.  Preamplifiers and the data acquisition hardware have been designed and tested. The water-cooled $^7$LiF target, mounted on an Al foil  is in a thin carbon fiber section of the beamline. The experiment will focus at first on a measurement of the internal pair creation from the 18.15 MeV state of $^8$Be.
%with the aim of searching for the X17 particle reported by the ATOMKI experiment and for other effects. 
Assuming the ATOMKI evaluation of the electron-pair  production rate from X17, a Geant4 simulation predicts observation of a clear signal after about two weeks of data taking with a 2 $\mu$A proton beam. The IPC measurement could eventually be extended to the giant dipole resonance of $^8$Be, as well as to other nuclei, in particular to $^{10}$B.
\end{abstract}

\section{Introduction}
 The Hungarian ATOMKI experiment~\cite{Krasznahorkay:2015iga,Krasznahorkay:2019lyl}  claimed observation, with a significance of more than 6 standard deviations, of a particle of mass around 17 MeV/c$^2$ in the spectrum of internal pair creation (IPC) from the decay of the 18.15 MeV state of $^8$Be, as well as in the decay of states of $^4$He near 20 MeV of excitation. This has aroused enormous interest and speculation on the nature of this so-called X17 particle. One description, consistent with present observations, would be a protophobic vector particle~\cite{Feng:2016ysn} for which the production rate depends principally, when off-resonance, on the E1 transition from direct proton capture~\cite{Zhang:2020ukq,Sas:2022pgm} .

 This contribution gives an update on the preparations of a new  experiment, being set up in Montreal, for an independent measurement of IPC from $^8$Be.
 %, confirming the existence of X17.  
 The Universit\'e de Montr\'eal 6 MV tandem Van de Graaff facility can deliver proton beams of 2$\mu$A (possibly up to 10 $\mu$A) with an energy resolution of 2 keV for proton energies in the range 0.4-1.0 MeV.
 The 18.15 MeV 1$^+$ state of $^8$Be is above the proton separation energy and decays principally by proton emission. Nevertheless, there is a small fraction of M1 electromagnetic decay to the 0$^+$ ground state. The IPC coefficient is 3.9 $\times 10^{-3}$ and the ratio of 
 branching ratios $B(X\to e^+e^-)/B(^8Be^*\to \gamma)$ is 5.8$\times 10^{-6}$, according to the  ATOMKI measurement. The production rate of the X17 particle is therefore extremely small. Nevertheless, it could, in principle,  be measured with a high significance since the IPC is dominated by low invariant mass of the $e^+e^-$ pair, and therefore a large boost and small opening angle, whereas the decay of the relatively heavy X17 particle would give a small boost and large opening angle to the $e^+e^-$ pair. The observation of a putative X17 requires, therefore, an experimental setup with large solid angle  acceptance, efficient rejection of gamma-ray background, and a precise  measurement of the directions of the electron and positron.
 
 \section{Experimental setup}
\begin{figure}[h]
%\begin{minipage}{20pc}
\includegraphics[width= 0.9\textwidth]{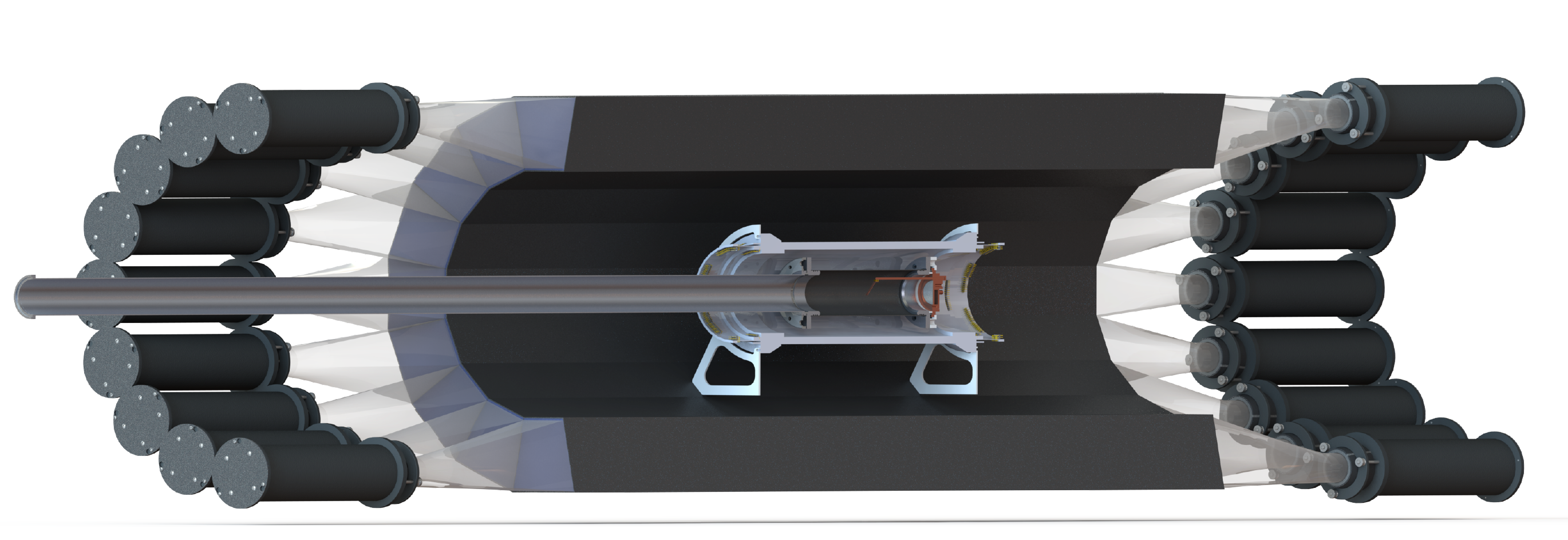}
%\caption{\label{fig:setup}Figure caption for first of two sided figures.}
%\end{minipage}\hspace{2pc}%
%\begin{minipage}{10pc}
\begin{center}
\includegraphics[width= 0.3\textwidth]{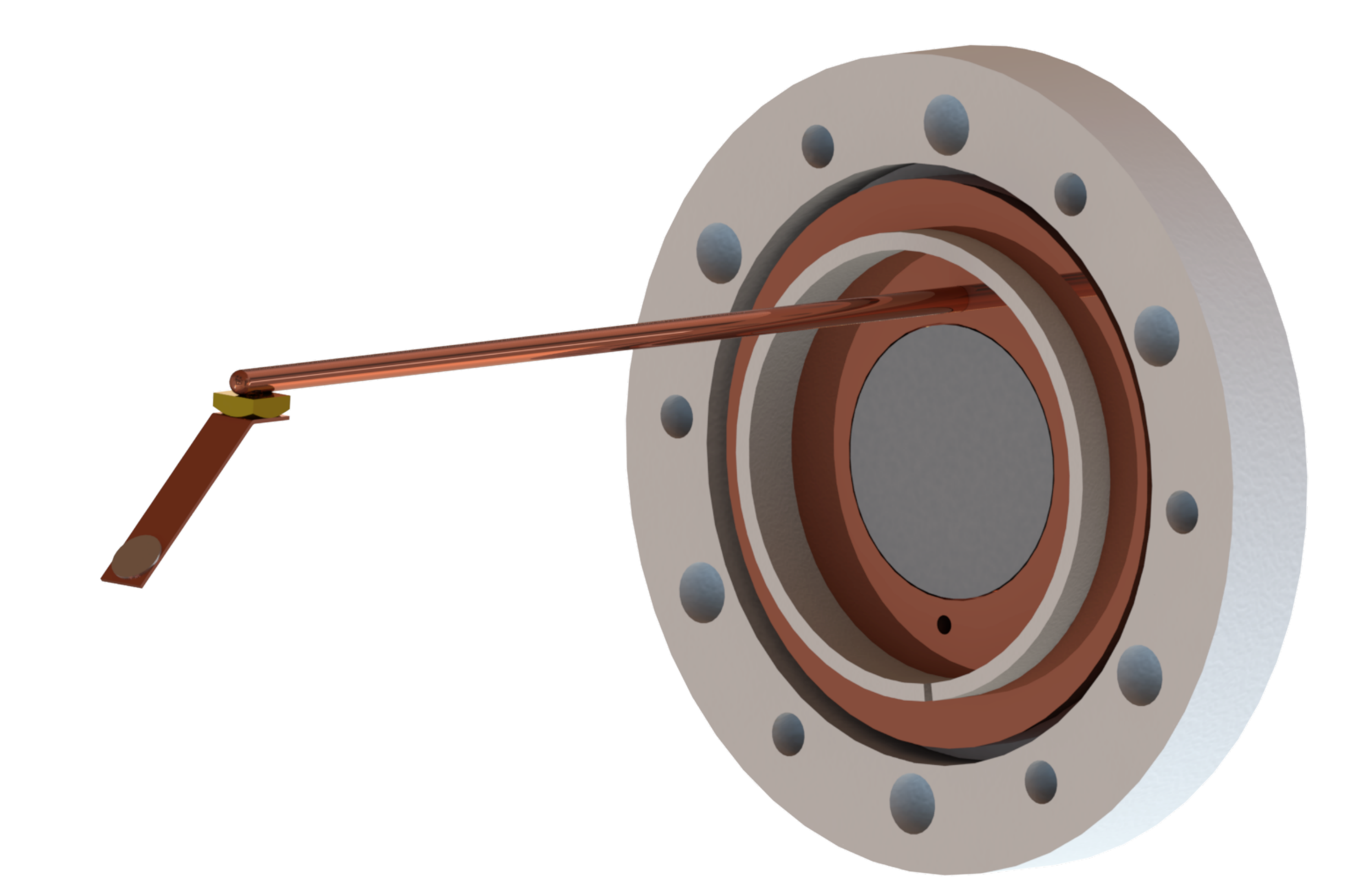}
\end{center}
\caption{\label{fig:setup}Basic experimental setup. Top: wire chamber surrounded by scintillator bars. Bottom: the target holder.}
%\end{minipage} 
\end{figure}

The experimental setup (Fig.~\ref{fig:setup}) consists principally of a cylindrical multiwire proportional chamber (MWPC) of inner radius 6 cm and a length of 36 cm, filled with 74\% Ar and 26\% CO$_2$ and surrounded by 16 scintillator bars of length 1.4 m arranged cylindrically at an inner radius of 16 cm. The chamber and the scintillator bars were previously used by the DAPHNE experiment~\cite{Audit:1991gq} at Saclay and Mainz. A number of phototubes are borrowed from TRIUMF. 
With its 192 anode wires with 2 mm spacing and 68 cathode strips of width 4 mm arranged at 45$^o$ to the anode wires, the angular resolution of the MWPC is around 2$^o$ FWHM assuming a 2 mm diameter beam spot. The chamber walls are made up of light material (1 mm of Rohacell) to reduce the effect of external pair creation (EPC) from gamma rays. Signals from the two ends of the scintillator bars allow a good measurement of the energy deposit of the electrons with an estimated resolution of 7\%/$\sqrt{E(MeV)}$ as well as a position resolution of $\sim 5$ cm FWHM along the bar.
The active detector volume covers a solid angle of 95\% of 4$\pi$. A second, larger, MWPC is available, but will not be used in the first stage.
A $^7$LiF target will be evaporated on a thin Al foil which will be inclined at 45$^o$ to the beam line and held by a  cooling rod. In the region of the target, the beamline is made of low density carbon fiber with a wall thickness of 0.8 mm. COMSOL simulations predict that the target temperature will be below 70$^o$, assuming 20 W heating from a proton beam of 20$\mu$A.

\begin{figure}[h]
\includegraphics[width= 0.22\textwidth]{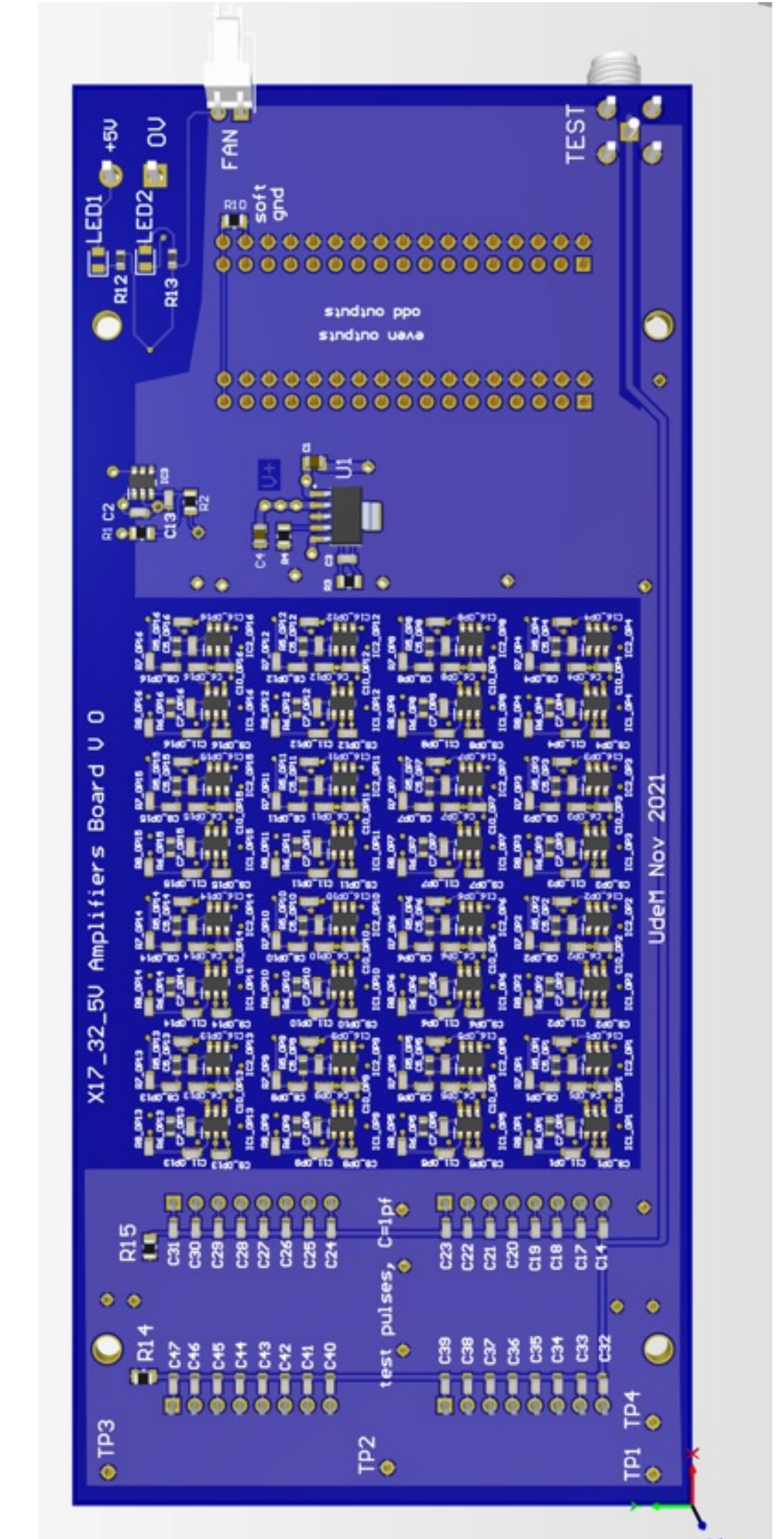}
\includegraphics[width= 0.78\textwidth]{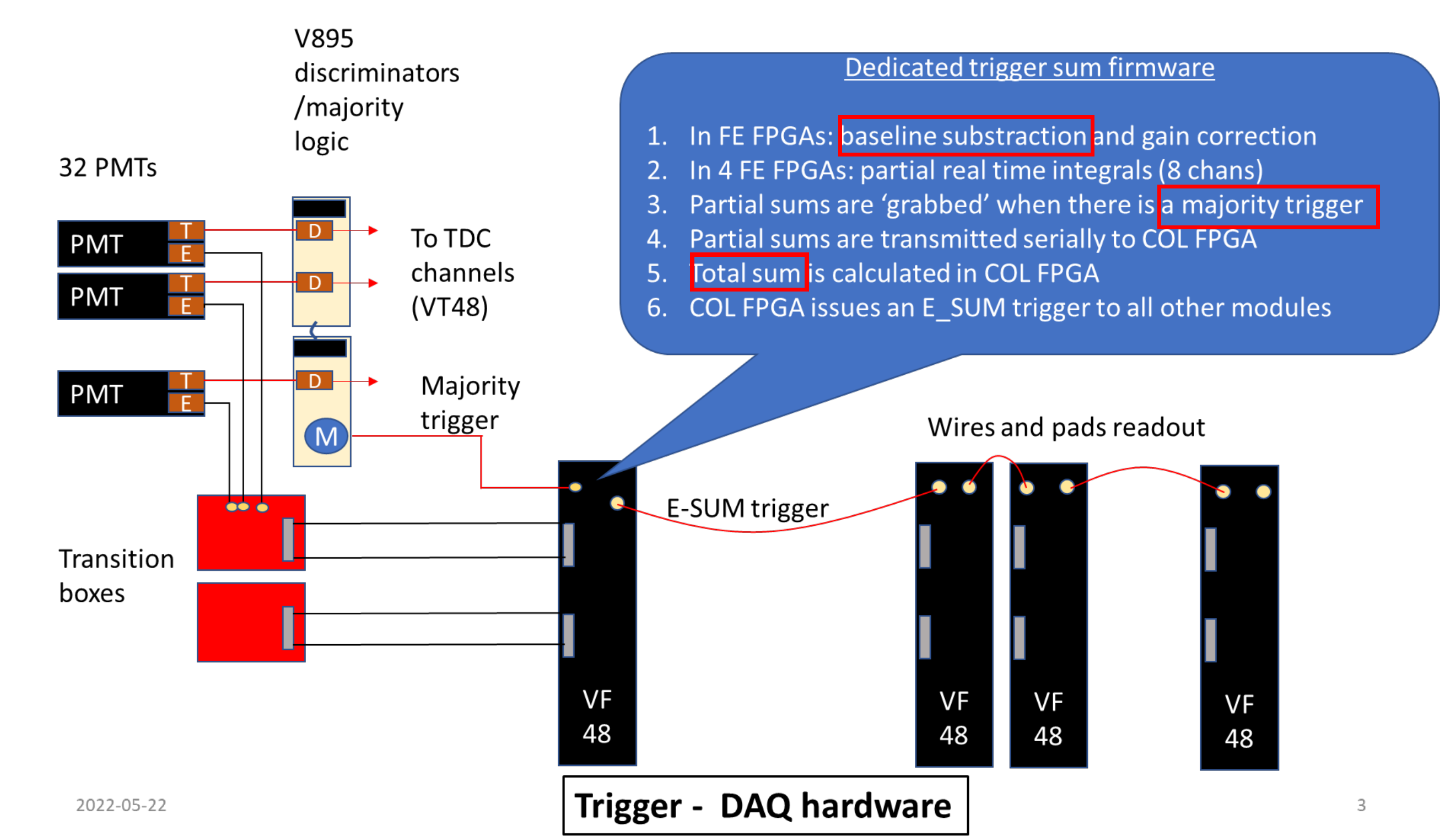}
\caption{\label{fig:trigger} Charge preamplifier (left)  and Level 1 trigger scheme (right).}
%\end{minipage} 
\end{figure}

Preamplifiers for the MWPC wires and strips~(Fig.~\ref{fig:trigger})  have been designed and tested. They will be placed close to the ends of the chamber. The signals are sent to VF48 ADC's and VT48 TDC's, designed in Montreal for an earlier experiment. The trigger~(Fig.~\ref{fig:trigger}) is based on a coincidence of signals from at least 2 scintillator bars and two hits from the MWPC wires. Front end FPGA's perform baseline subtraction and gain correction for the photomultiplier signals. The sum of energy depositions in the scintillator bars is calculated and a threshold is set for the trigger. The firmware for the data acquisition has been tested with a partial number of channels.

 \section{Monte Carlo simulation}
\begin{figure}[h]
\includegraphics[width= 0.45\textwidth]{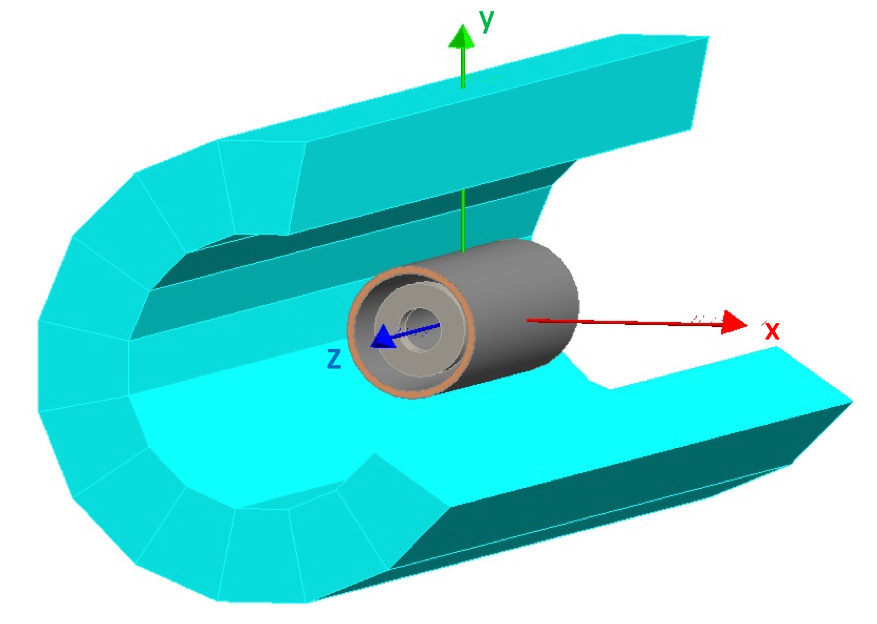}
\includegraphics[width= 0.55\textwidth]{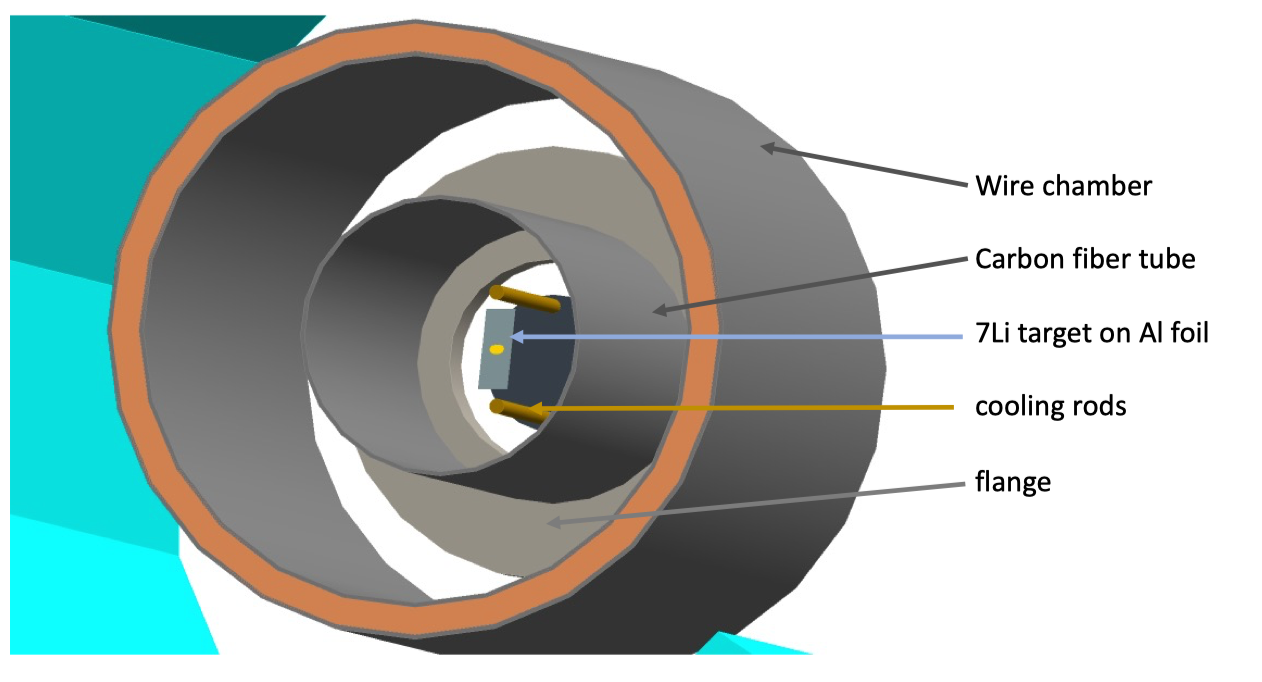}
\caption{\label{fig:MCgeometry} Geometry of the detector in Geant4 Monte Carlo.}
\end{figure}

 The Monte Carlo simulation  of the experimental setup with GEANT4 is well advanced. The basic elements of the detector are included (see Fig.~\ref{fig:MCgeometry}) and the resolutions of the energy measurements in the scintillators and of the electron directions in the MWPC are taken into account. 
 Thanks to the large solid angle acceptance of the experimental setup, the reconstruction efficiency will have a smooth dependence on the electron pair opening angle. This would be an improvement over the ATOMKI setup, which consists of 6 isolated detectors~\cite{Gulyas:2015mia}
 
 The IPC spectrum from M1 as well as from E1 transitions are simulated, using expressions from Rose~\cite{Rose:1949zz} and Stiebing~\cite{Stiebing:2003zs}. 
The reconstruction of the X17 signal will adopt the strategy of the ATOMKI experiment. It is principally based on the measurement of the opening angle between the $e^+$ and the $e^-$ since it is directly correlated with the invariant mass of the pair. For a low invariant mass, which is where most of the IPC spectrum lies, large opening angles are also correlated with a large asymmetry between the  $e^+$ and the $e^-$  energies. A selection cut on this asymmetry is therefore also applied: $y \equiv |E_1-E_2|/(E_1+E_2) $. Finally, to reduce background from poorly reconstructed electron pairs, an invariant mass selection will also be applied. A machine learning algorithm is being considered for selection of signal over background.
 
\begin{figure}[h]
\includegraphics[width= 0.5\textwidth]{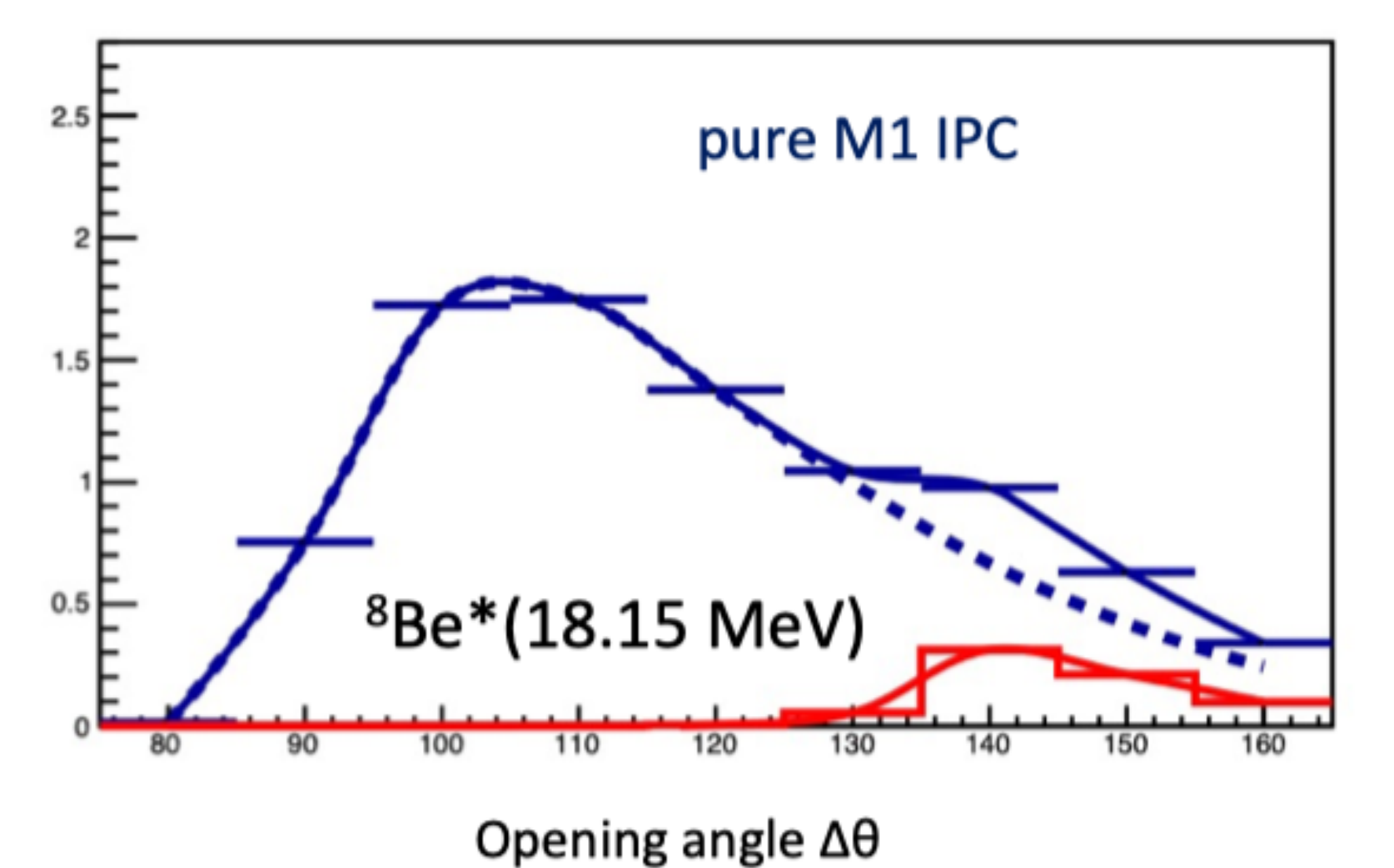}
\includegraphics[width= 0.5\textwidth]{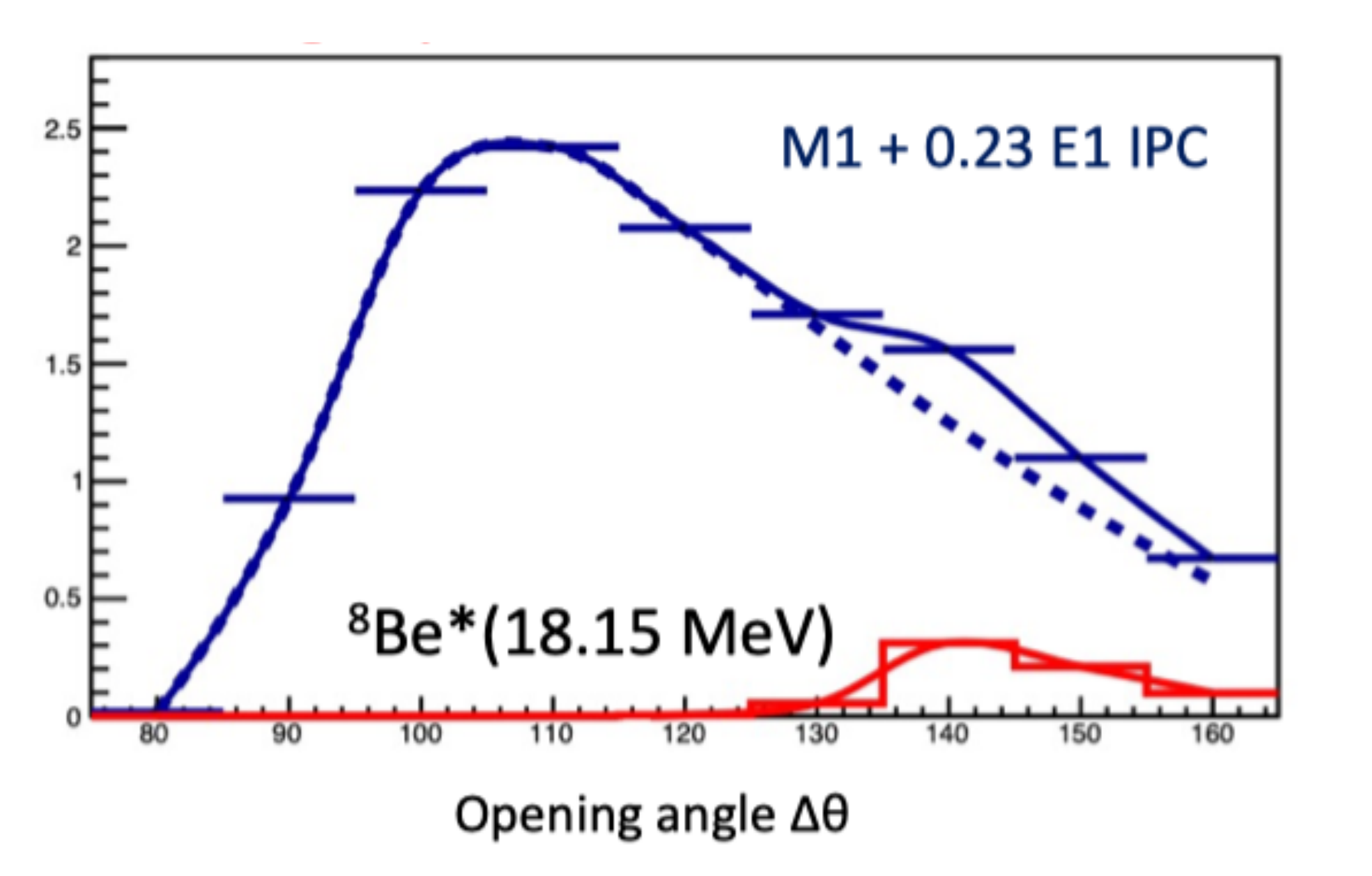}
\caption{\label{fig:angle} Predicted opening angle distributions (see text for details).}
%\end{minipage} 
\end{figure}

  Based on the measured rate of the 478 keV photons from the $^7Li(p,p'\gamma)^7Li$  reaction in beam tests with BGO and Ge(Li) detectors, and by extrapolation to 95\% of $4\pi$ acceptance, and assuming the IPC to gamma ratio above, it is estimated that the rate of X17 production would be $\sim 9 $ h$^{-1}$ with a 2$\mu$A proton beam if the ATOMKI measurements were confirmed. Under these conditions, the rate of the IPC background, in the region of interest corresponding to an opening angle of the $e^+e^-$ pair of around 140$^o$, is expected to be about 15 h$^{-1}$. The predicted opening angle distributions, based on the X17/IPC ratio and assuming a pure M1 or a mixture of M1 and E1 transitions, are shown in Fig.~\ref{fig:angle}. The background depends very sensitively on the ratio of E1/M1~\cite{Sas:2022pgm} and will need to be fitted in a control region outside the expected excess.   For these plots, the asymmetry parameter is limited to $y < 0.45$ and the invariant mass $m_{ee}$ is greater than 12 MeV/c$^2$. The distributions do not account for statistical fluctuations. They are  normalized to one X17 particle at the ATOMKI rate and assume good calibration of the energy depositions in the different scintillators. Clearly, with 1-2 weeks of data taking, with a X17 production of 9 h$^{-1}$, the signal should be very significant if it exists.
 
\section{ Conclusion and outlook}
An experimental setup for an independent evaluation of the purported existence of the X17 particle is in preparation in Montreal. All the elements are being put in place: target, beamline, MWPC, scintillator bars. The electronic data acquisition and readout are nearly ready.  Beam tests are expected to start in autumn 2022, followed by calibrations and tuning. Data taking should take place in early 2023. The large solid angle acceptance will allow a clear measurement of the opening angle of the electron-positron pairs. It will also help in defining the quantum numbers of the X17 if it exists as the distribution, for different polar angles of the electrons,  depends on the nature of the particle~\cite{Viviani:2021stx}.

As mentioned in the introduction, IPC in highly excited states of other nuclei can help to evaluate the existence of X17.
%and understand its nature.
Among other possibilities being considered are the E1 transitions from the  19.3 MeV state of $^{10}B$, produced with the reaction $^7Li(^3He,\gamma)^{10}B$. Compared to the $^8Be^*$ transition, because of the higher energy, the comparable production cross section, and the E1 nature of the transition, the signal can be significantly more prominent. Also, since X17 production off-resonance is dominated by direct proton capture, it is expected to be strongly enhanced at higher proton energies populating the giant dipole resonance of $^8Be^*$~\cite{Fisher:1976zza}.

\section*{References}

%Alternatively uncomment the line below to use a .bib file 
%\bibliography{iopart-num.bib}

\end{document}